\documentstyle[12pt]{article}
\begin{document}

\begin{titlepage}
\vspace*{-4cm}
\begin{flushright}
EFI-95-50 \\
hep-ph/9508322
\end{flushright}
\vspace{1cm}
\begin{center}
\Huge Fermion Masses and Mixings \\
\Huge in a String-Inspired Model
\end{center}
\vspace{1cm}
\begin{center}
\large Merle Michael Robinson and Jacek Ziabicki
\end{center}
\begin{center}
\normalsize\it Enrico Fermi Institute and
the Department of Physics,\\
\normalsize\it University of Chicago, 5640 S. Ellis Ave.,
Chicago, IL 60637
\end{center}
\vspace{1cm}
\vspace{1cm}

\begin{abstract}

In this paper we study quark and lepton mass matrix textures
in a model containing an additional $U(1)_X$ gauge symmetry
with origins in string compactification. The $U(1)_X$ symmetry
is broken near the string scale, and we assume that the anomalies
are canceled by the Green-Schwarz mechanism. We also assume that
fermion mass matrices are generated by an additional scalar field
through an approach analogous to that of Froggatt and Nielsen.
By requiring that supersymmetry not be broken at the high scale,
we can derive the vacuum expectation value of this scalar field to
then predict fermion masses and mixings for any given $X$ charge
assignment. We examine the possible solutions, and although in the
simplest model they do not completely agree with experiment, the
results are close enough to merit further inspection.

\end{abstract}

\end{titlepage}

\section{Introduction}

Understanding the fermion mass structure has been a goal of
particle theorists for some time. In 1978, Froggatt and Nielsen
\cite{fn} found that a spontaneously broken family dependent
symmetry could naturally explain the large mass ratios among
different families of quarks and leptons. Renormalizing
experimental data to the Planck scale reveals the order of
magnitude estimates to the following ratios \cite{op,rr,br}:
\begin{eqnarray}
{m_u \over m_t} = {\cal O}(\lambda^8)\;; \qquad\qquad
{m_d \over m_b}&=&{\cal O}(\lambda^4)\;; \qquad\qquad
{m_e \over m_\tau}={\cal O}(\lambda^4) \nonumber\\
{m_c \over m_t} = {\cal O}(\lambda^4)\;; \qquad\qquad
{m_s \over m_b}&=&{\cal O}(\lambda^2)\;; \qquad\qquad
{m_\mu \over m_\tau} = {\cal O}(\lambda^2)\;,
\label{eqratios}
\end{eqnarray}
where $\lambda\simeq 0.22$ is the small parameter used in the
Wolfenstein's pa\-ra\-me\-tri\-za\-tion \cite{w} of the
Cabibbo-Kobayashi-Maskawa (CKM) mixing matrix \cite{c,km}.

In the past few years, there has been a revival of theories which
predict a mass hierarchy from a spontaneously broken family
symmetry. This time the work has been done in the context of
supersymmetry (SUSY). The general idea has been used widely in
more detailed models with family symmetries that were continuous
and discrete, Abelian and non-Abelian, global and local, and with
different choices for the symmetry breaking scale
\cite{br,ns,ln,ks,ir,dp}.

One of the unanswered questions of the original Froggatt-Nielsen
model is the origin of the family symmetry breaking.  It has been
suggested \cite{br,ir} that the supersymmetric versions of the
model may be derived from superstring compactification, where
spontaneously broken anomalous $U(1)$ gauge symmetries typically
occur. In models where the anomalies are canceled by the
Green-Schwarz mechanism \cite{gs}, the symmetry breaking scale is
slightly below the string scale.  Preserving supersymmetry at the
high scale determines the vacuum expectation value (VEV) of the
symmetry breaking field $\theta$ and the hierarchy parameter
$\lambda$, which greatly restricts the theory.  The purpose of our
work is to find out whether this new constraint can be accommodated
in a phenomenologically acceptable model.

In this paper, we present models which predict the fermion masses
and mixings in a string-inspired framework. Since we do not work
with an exact string model, we carry a model-independent analysis
as far as possible. In doing so we make the following assumptions:
(1) the additional Froggatt-Nielsen symmetry is an anomalous $U(1)$
originating in string theory so that the anomalies are canceled by
the Green-Schwarz mechanism;
(2) renormalization of couplings and particle masses is done within
the framework of the minimal supersymmetric standard model (MSSM);
(3) the $U(1)$ symmetry is broken by the VEV of only one field,
$\theta$;
(4) the Yukawa coupling of $\theta$ to the fermions $f_\theta$ is
one.
For most of the paper, we assume that the Kac-Moody level,
$k_{GUT}$, for the Grand Unified Theory (GUT) group is one, but we
keep $k_{GUT}$ as a parameter in most equations. Finally, in the
context of a particular string model, the $U(1)$ symmetry breaking
field(s) will be known, as will the value of $f_\theta$ and
$k_{GUT}$.

The outline of the paper is as follows. In section~\ref{secreview},
we summarize the main features of the Froggatt-Nielsen models. In
section~\ref{secsusy} we discuss the implications of unbroken
supersymmetry for the value of the mass hierarchy parameter,
$\lambda$. This is followed by section~\ref{secgsmech}, which gives
the background and some important facts about the Green-Schwarz
mechanism. Section~\ref{secanomaly} is dedicated to the anomalies
of the model. We show there how $\lambda$ depends only on the $X$
charges of the standard model fields. In section~\ref{secpq} we
present some constraints on those charges following from relations
(\ref{eqratios}). We illustrate these constraints in
section~\ref{secexamples} by working out in detail a few simple
models. Section~\ref{secconclusions} summarizes our results.

\section{Froggatt-Nielsen Models}
\label{secreview}

\begin{figure}
\begin{center}
\newsavebox\dashed
\begin{picture}(300,100)
\put(5,0){\line(1,0){290}}
\sbox{\dashed}{\multiput(0,0)(0,10){8}{\line(0,1){5}}%
\put(-4.58,72){$\times$}\put(-7.5,85){$\langle\theta\rangle$}}
\multiput(50,0)(0,10){8}{\line(0,1){5}}\put(50,80){\line(0,1){3}}
\put(45,85){$H$}
\multiput(90,0)(40,0){5}{\usebox\dashed}
\multiput(30,0)(40,0){7}{\vector(1,0){0}}
\put(20,10){$+2$}
\put(60,10){$+2$}
\put(100,10){$+1$}
\put(147,10){0}
\put(180,10){$-1$}
\put(220,10){$-2$}
\put(260,10){$-3$}
\end{picture}
\end{center}
\caption{Tree diagram leading to effective Yukawa interactions.
Above the solid line are $U(1)$ charges of quarks and mirror
quarks.}
\label{figfeynman1}
\end{figure}
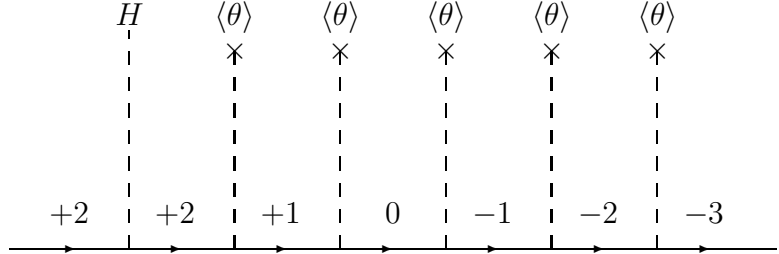

Originally, Froggatt and Nielsen proposed \cite{fn} that a
flavor-dependent symmetry be broken by the VEV of an additional
scalar field, $\theta$, which would be a singlet of the standard
model gauge groups. Their idea also assumed a set of heavy ``mirror
quarks'', analogous to the standard model quarks, with a spectrum
of charges under the horizontal symmetry.  Mass matrices would then
arise through effective Yukawa interactions resulting from Feynman
diagrams such as that in figure~\ref{figfeynman1}.%

In figure~\ref{figfeynman1}, we show an example, where the $X$
charge assignments of the quarks are written above each quark line.
On one side of the diagram, we have a left-handed quark doublet
with charge $+2$. If we assign a charge of $-3$ to the right-handed
quark, then there must be five $\theta$ interactions, with five
mirror quarks of charges $+2$, $+1$, 0, $-1$, $-2$ in between. The
first mirror quark of charge $+2$ interacts with the standard model
Higgs doublet and the quark doublet to conserve $SU(2)$ symmetry.
Assuming a common mass $M$ for the mirror quarks, and a common
Yukawa coupling $f_\theta$ of the $\theta$ field to all the quarks,
$\langle\theta\rangle$ should take a value such that
$\lambda \sim f_\theta \langle\theta\rangle/M$. The mass term
resulting from fig.~\ref{figfeynman1} would be
\begin{equation}
\label{eqquarka}
f_u \bar u_j Q_i H
\left(f_\theta {\langle\theta\rangle\over M} \right)^5.
\end{equation}
In a realistic model, the charges must be assigned so that all the
mass matrix eigenvalues agree with the relations~(\ref{eqratios})
above.

Here, we do not make the Froggatt-Nielsen assumptions, and thus do
not require mirror quarks. We assume that the flavor symmetry is a
gauged $U(1)$ symmetry, labeled by $X$, left over from string
compactification. We expect the action to contain all terms
consistent with charge conservation. Such terms appear due to
string tree diagrams; therefore, the effective Yukawa coupling
$f_\theta$ will be a product of the string coupling constant
$g_s$ and other terms of order unity. Not knowing the details of
the model, we cannot evaluate $f_\theta$, and here assume
$f_\theta=1$. In order to demonstrate the generation of mass
terms, we give an explicit example. We do not make any further
assumptions about the physical mechanism.

First, we define the $X$ charges to be $q_{Qi}$, $q_{ui}$,
$q_{di}$, $q_{Li}$, and $q_{ei}$ for the left-handed quark
doublets, the left-handed up-type antiquarks, the left-handed
down-type antiquarks, the left-handed lepton doublets and the
left-handed positrons ($i$\/ is the family index). Also, we define
$q_H$ to be the sum of the $X$ charges for the two Higgs doublets
of the supersymmetric standard model. We then consider all
bilinear fermion terms that conserve $X$ charge. A typical up-type
quark term would be
\begin{equation}
\label{eqquark}
\left({\bf Y}_u\right)_{ij}=
f_u \bar u_j Q_i H_1
\left({\langle\theta\rangle\over M}\right)^{q_{Qi}+q_{uj}+q_{H1}}
\end{equation}
The entire Yukawa mass matrix then follows:
\begin{equation}
\label{eqmatrix}
{\bf Y}_u=
f_u \,\lambda^{q_{H1}}
\left(
\begin{array}{ccc}
\lambda^{q_{Q1}+q_{u1}}
&\lambda^{q_{Q1}+q_{u2}}
&\lambda^{q_{Q1}+q_{u3}}\\
\lambda^{q_{Q2}+q_{u1}}
&\lambda^{q_{Q2}+q_{u2}}
&\lambda^{q_{Q2}+q_{u3}}\\
\lambda^{q_{Q3}+q_{u1}}
&\lambda^{q_{Q3}+q_{u2}}
&\lambda^{q_{Q3}+q_{u3}}
\end{array}
\right).
\end{equation}

\section{Implications of Unbroken Supersymmetry}
\label{secsusy}

The basic premise of this work is the assumption of a deeper
connection between string theory and supersymmetric models with
spontaneously broken family symmetries. By assuming such a
connection, we can ``borrow'' a $U(1)$ gauge symmetry left over
from string compactification.

We begin with $N=1$ global supersymmetry and the scalar potential:
\begin{equation}
V={1\over2}\sum_\alpha (D^\alpha)^2 + \sum_i |F_i|^2.
\end{equation}
Here $F_i = \partial W/\partial\phi_i$. We do not specify the
superpotential $W$; thus, we will not be able to predict the VEV's
of all of the fields in the model. The gauge $D$ term is given by
$D^\alpha=g_{(\alpha)}\sum_{ij}\phi_i^\ast(T^\alpha)_{ij}\phi_j$,
where $\phi_i$ are the matter chiral superfields, $T^\alpha$ are
the generators of the gauge group, and $g_{(\alpha)}$ are the gauge
couplings. For an anomalous $U(1)$ gauge group, the corresponding
$D$ term will be modified by a Fayet-Iliopoulos term. Its
magnitude has been calculated in string theory \cite{ad,di,ds} on
the assumption that the anomalies are canceled by the
Green-Schwarz mechanism \cite{gs}:
\begin{equation}
\label{eqDterm}
D={g_s M_s^2\over192\pi^2} {\,\rm tr\,} Q + \sum_i q_i |\phi_i|^2,
\end{equation}
where $g_s$ is the renormalized string coupling constant, $M_s$ is
the string scale, and ${\,\rm tr\,} Q \equiv \sum_i q_i$ is the sum of the
$U(1)$ charges of all the particles. (See also \cite{fi} for a clear
exposition.) In the model we are studying,
the $U(1)_X$ family symmetry gauge group is anomalous, has origins
in string theory and requires a $D$ term given by
eq.~(\ref{eqDterm}).

The supersymmetric vacuum requires $\langle F_i\rangle=0$ and
$\langle D^\alpha\rangle=0$. If the $X$ charges of all the
particles are of the same sign, then, according to
eq.~(\ref{eqDterm}), it is impossible to preserve supersymmetry. We
therefore require that charges with both signs be present.
(Fortunately, this is typically the case in string
compactifications.) As a convention, we give the $\theta$ field a
negative charge. In section~\ref{secexamples}, we will consider
both the case in which all the standard model matter fields have
positive $X$ charge and the case in which they can have charges of
either sign, subject to the condition ${\,\rm tr\,} Q>0$.

We have to assume that none of the fields charged under the
standard model (SM) gauge groups can develop vacuum expectation
values---otherwise color or electroweak symmetry would be broken at
the high scale $\sim M_s$. The problem of flat directions here
does not differ from the problem of flat directions in the MSSM.
We require $\langle D^\alpha\rangle=0$ for each gauge factor
separately; for each of the SM gauge groups this condition involves
only SM fields, so that the flat directions will be the same as in
the MSSM. For $U(1)_X$, setting $\langle D\rangle=0$ merely
determines $\langle\theta\rangle$ and does not constrain MSSM
fields. Hence, just as in the MSSM, we have to rely on the
superpotential to lift the flat directions.

The most important implication of eq.~(\ref{eqDterm}) for this work
is that preserving supersymmetry determines the VEV of the $\theta$
field
\begin{equation}
\label{eqsqrt}
{\langle\theta\rangle\over M_s} =
\sqrt{{g_s\over192\pi^2} {{\,\rm tr\,} Q \over |q_\theta|}}\;.
\end{equation}
If we assign $X$ charges to all the fields and use $\theta$ as the
family symmetry breaking field in the Froggatt-Nielsen scheme, we
obtain a prediction for the Yukawa mass matrices.

The question we attempt to answer in this paper is: can we find a
set of charges for all the standard model fields, consistent with
the requirements of anomaly cancellation, that will predict
phenomenologically viable powers of $\lambda$ (much like the work
by Ib\'a\~nez and Ross \cite{ir}, but without the assumption of
left-right symmetry) \ {\em and\/} \ predict a phenomenologically
viable value of $\lambda\simeq0.22\,$? We shall see that this is
possible as long as $f_\theta$ is not very different from
unity.

Our results, as can be seen from eq.~(\ref{eqsqrt}), are rather
sensitive to the value of the string coupling constant, $g_s$. At
the unification scale, we use the tree-level relation \cite{k,g}
\begin{equation}
\label{eqkgut}
1/g_s^2=k_{GUT}/g_{GUT}^2,
\end{equation}
where $k_{GUT}$ is the Kac-Moody level for corresponding GUT gauge
group algebra. Here, we take $k_{GUT}=1$. We then use a typical
value $\alpha_{GUT} = {g_{GUT}}^2/4\pi \simeq 1/25$ to get
$g_s \simeq 0.7$.  Substituting into eq.~(\ref{eqsqrt}) with
$q_\theta=-1$, we obtain
\begin{equation}
\lambda = \langle\theta\rangle/M_s =
1.92 \times 10^{-2} \, \sqrt{{\,\rm tr\,} Q}.
\end{equation}

\section{Green-Schwarz Anomaly Cancellation}
\label{secgsmech}

In chiral theories, it is necessary to consider the problem of
quantum anomalies. These anomalies to classical symmetries are
dangerous in that they prevent the existence of gauge theories. In
this section we discuss a method of removing anomalies that may
arise with a new $U(1)_X$ gauge symmetry.

We start with $U(1)$ chiral transformations on all fermions:
\begin{eqnarray}
\label{eqtransf}
\Psi(x)&\to& \exp\left[-iq\gamma_5 \Theta(x)\right]\Psi(x)\\
\bar\Psi(x)&\to& \bar\Psi(x)
\exp\left[-iq\gamma_5 \Theta(x)\right],
\nonumber
\end{eqnarray}
where $q$ is the charge of each fermion. Since the path integral
measure is not invariant under the transformation, we obtain
new terms beyond the usual current divergence \cite{f}.
The difference can be expressed as a change in the Lagrangian:
\begin{equation}
\label{eqLag}
{\cal L}\to {\cal L}
- \Theta(x)\!\!\!\!\sum_{i=1,2,3,X}\!\!\!\! C_i F_i \tilde F_i
- \Theta(x) \left(\partial_\mu j_5^\mu\right)
\end{equation}
where the sum is over the standard model gauge groups and
$U(1)_X$. The coefficient $C_1 = {\,\rm tr\,}\left[Q (Y/2)^2\right]$ is the
mixed $U(1)_X \left(U(1)_Y\right)^2$ anomaly, $C_X = {\,\rm tr\,} Q^3$ is
the $\left(U(1)_X\right)^3$ anomaly, and
$C_{2,3} = {1\over2} {\rm tr}_{2,3}\,Q$ are the
$U(1)_X \left(SU(2)_L\right)^2$ and $U(1)_X \left(SU(3)_c\right)^2$
anomalies. (The trace $\,{\rm tr}_{2,3}\,Q$ is over fermions with
$SU(2)_L$ and $SU(3)_c$ charge, respectively.) In some cases we can
choose the charges so that all the anomaly coefficients are zero,
but here we examine the possibility of canceling the anomalous
term with another of opposite sign \cite{gs}. There is another
mixed anomaly with $U(1)_Y$, the $(U(1)_X)^2\, U(1)_Y$, but it does
not fit into the discussion. It results from a different
transformation than that in eq.~(\ref{eqtransf}), where $q$ would
be the $U(1)_Y$ instead of the $U(1)_X$ charge.

In the Green-Schwarz mechanism, anomalies are canceled through an
additional field. In string theory, the antisymmetric tensor
$B_{\mu\nu}$ naturally serves this purpose. In four dimensions
\cite{hn}, we can replace $H=dB$ with its dual, which is the
derivative of the axion field $\Phi$:
\begin{equation}
\label{eqdual}
dB = *d\Phi.
\end{equation}
This field couples to the gauge groups in the following way:
\begin{equation}
\label{eqetacoupl}
\Phi\!\!\!\!\sum_{i=1,2,3,X}\!\!\!\! k_i F_i \tilde F_i,
\end{equation}
where $k_i$ are the Kac-Moody levels of the corresponding gauge
algebra. For the $U(1)_X$ gauge transformation,
\begin{equation}
\label{eqXgauge}
A_X^\mu \rightarrow A_X^\mu + \partial^\mu\Theta(x)
\end{equation}
$\Phi$ follows the transformation
\begin{equation}
\label{eqetagauge}
\Phi\rightarrow\Phi + \Theta(x) \,\delta_{GS}.
\end{equation}
Therefore, we can remove quantum anomalies through a gauge
transformation if $\delta_{GS}=C_1/k_1=C_2/k_2=C_3/k_3=C_X/k_X$.
For a more detailed discussion, we refer the reader to the paper by
Ib\'a\~nez \cite{i}.

So far, we have ignored gravity, but the conclusions do not change.
We must only cancel one additional anomaly,
$C_{\rm grav} R \tilde R$, through a gauge transformation on one
additional coupling $k_{\rm grav} \Phi R \tilde R$. Finally, we
have:
\begin{equation}
\label{eqckratio}
C_1/k_1=C_2/k_2=C_3/k_3=C_X/k_X=C_{\rm grav}/k_{\rm grav}.
\end{equation}
In this paper, we take $k_1=5/3$ and $k_2=k_3=k_{\rm grav}=1$,
and we do not use $k_X$, so that
\begin{equation}
\label{eqcratio}
C_1:C_2:C_3:C_{\rm grav} = {5\over3}:1:1:1.
\end{equation}

\section{Quantum Anomalies}
\label{secanomaly}

We can now apply the results of the previous section to
the case at hand. The mixed anomalies with the standard
model gauge groups are
\begin{eqnarray}
\label{eqC123}
C_1 &=& {1\over6} (q_Q + 8q_u + 2q_d +3q_L +6q_e +3q_H)
\nonumber\\
C_2 &=& {1\over2} (3q_Q + q_L + q_H) \\
C_3 &=& {1\over2} (2q_Q + q_u + q_d), \nonumber
\end{eqnarray}
where $q_Q = \sum_{i=1}^3 q_{Qi}$, etc., but
$q_H = q_{H1}+q_{H2}$ is the
sum of the $U(1)_X$ charges of the two Higgs doublets.
Our calculations do not include the right-handed neutrinos,
but since we allow for the existence of additional particles
with $X$ charge that are singlets under the standard model,
our results do not depend on the existence and $X$ charge
assignments of $\nu_R$.

The gravitational anomaly is given by
\begin{equation}
\label{eqcgrav}
C_{\rm grav} = {1\over24}
\sum_{\mbox{\scriptsize all particles}}\!\!\!\!\!\! q_i =
{1\over24}
(6q_Q + 3q_u + 3q_d + 2q_L + q_e + 2q_H + q_\theta + q_X),
\end{equation}
where $q_X$ is the sum of the $U(1)_X$ charges of any
additional fields which are singlets under the standard
model. We are not excluding such fields, and we cannot
evaluate $C_{\rm grav}$ directly. However, because we are
using the Green-Schwarz mechanism, we know that
$C_{\rm grav}$ must be in the correct
proportion~(\ref{eqcratio}) to the other anomalies.
{}From the expressions (\ref{eqC123}) and
(\ref{eqcgrav}) we then obtain
\begin{equation}
\label{eq123g}
q_\theta + q_X = 18q_Q + 8q_u + 7q_d = 14C_3 + 4q_Q + q_u.
\end{equation}
With the assumption that $\theta$ is the only field with a negative
$U(1)_X$ charge, we see immediately that we {\em require\/} the
extra fields with no standard model interactions to balance
eq.~(\ref{eq123g}) with a large positive contribution $q_X$. If we
allow the quarks to have negative charges, this is no longer true.
Even then, in section~\ref{secpq} we shall see that $C_3 \sim 9$
in phenomenologically interesting models, so that the typical
model will require $q_X>0$.

The additional fields responsible for $q_X\not=0$ prevent us from
calculating the cubic anomaly $\left(U(1)_X\right)^3$, which then
does not impose any constraints on the model. We simply assume
that the charges of the extra fields are such that it is canceled:
\begin{equation}
C_X = \!\!\!\!
\sum_{\mbox{\scriptsize all particles}} \!\!\!\!\!\! q_i^3.
\end{equation}
On the other hand, the mixed anomaly
$\left(U(1)_X\right)^2 \, U(1)_Y$,
\begin{equation}
C_{YXX} = \!\!\!\!
\sum_{\mbox{\scriptsize all particles}} \!\!\!\!\!\! Y_i q_i^2
= \sum_{i=1}^3 \left(
q_{Qi}^2 - 2q_{ui}^2 + q_{di}^2 - q_{Li}^2 + q_{ei}^2
\right) - q_{H1}^2 + q_{H2}^2,
\end{equation}
depends only on the charges of the standard model particles and
cannot be canceled by the Green-Schwarz mechanism. For each
charge assignment we have to check that $C_{YXX}=0$.

The equality of $C_3$ and
$C_{\rm grav}$, eq.~(\ref{eqcratio}), is
crucial in this paper. Section~\ref{secsusy}, eq.~(\ref{eqsqrt})
gave us a prediction for the hierarchy parameter $\lambda$
in terms of ${\,\rm tr\,} Q$, the sum of the $U(1)_X$ charges of all the
particles in the model. That sum, according to
eq.~(\ref{eqcgrav}), is proportional to the gravitational
anomaly. Because $C_{\rm grav}=C_3$, every charge assignment
for the standard model fields (in fact, for the quark fields
alone) results in a prediction
\begin{equation}
\label{eqc3l}
\lambda = \sqrt{g_s/8\pi^2}\,\sqrt{C_3} \simeq 0.094\sqrt{C_3}.
\end{equation}

\section{Determinants of the Mass Matrices}
\label{secpq}

The product of the determinants of up- and down-quark mass matrices
will give us an important constraint. From eq.~(\ref{eqratios}), it
is immediately seen that
\begin{equation}
\label{eqlam}
\prod_{\mbox{\scriptsize all quarks}}
\!\!\!\!\! m_q \sim f_u^3 f_d^3\,\lambda^{18}.
\end{equation}
This should be compared to the product of the determinants of the
Yukawa matrices predicted by the model.
Writing ${\bf Y}_u$ in the form (\ref{eqmatrix}), we see that
every term in the determinant is of the order
$f_u^3 \,\lambda^{q_Q+q_u+3q_{H1}}$.
Similarly, every term in $\det {\bf Y}_d$ is of the order
$f_d^3 \,\lambda^{q_Q+q_d+3q_{H2}}$. The two taken together give
\cite{n}
\begin{equation}
\label{eqdet1}
\prod_{\mbox{\scriptsize all quarks}} \!\!\!\!\! m_q =
\left| \det{\bf Y}_u \right| \: \left| \det{\bf Y}_d \right|
\sim f_u^3 f_d^3 \,\lambda^{2q_Q + q_u + q_d + 3q_H}.
\end{equation}
Now, $q_H=0$ if a $\mu$ term $\mu H_1 H_2$ is to be allowed and
not suppressed. (For alternatives, see \cite{n,js}.
We note that a small change in $q_H$ can be easily accomodated,
as it will not change the predicted mass ratios or mixings.)
Then, using eq.~(\ref{eqC123}), we are left with
\begin{equation}
\label{eqdet2}
\prod_{\mbox{\scriptsize all quarks}} \!\!\!\!\! m_q
\sim f_u^3 f_d^3 \,\lambda^{2 C_3}.
\end{equation}
{}From equations (\ref{eqlam}) and (\ref{eqdet2}), we see that
\begin{equation}
\label{eqc318}
\lambda^{18} \sim \lambda^{2C_3},
\end{equation}
or
\begin{equation}
\label{eqc39}
C_3\simeq9,
\end{equation}
as was found earlier by Bin\'etruy and Ramond \cite{br}. This is
true whether or not there are texture zeros, provided that neither
determinant is zero. For $C_3=9$, eq.~(\ref{eqsqrt}) gives
$\lambda=0.28$.

The above reasoning assumes that $\lambda$ is fixed at about
$0.22$---the mass ratios (\ref{eqratios}) come from experiment,
not from assumptions about hierarchy. In this paper, we
derive the hierarchy parameter $\lambda$ from supersymmetry.
Taking $\lambda$ as predicted by eq.~(\ref{eqc3l}), we have to
replace (\ref{eqc318}) by
\begin{equation}
\label{eqc3f}
0.22^{18} \sim \left(0.094\sqrt{C_3}\right)^{2C_3}.
\end{equation}
Solving this, we get $C_3 \simeq 12.5$ and $\lambda\simeq0.33$,
rather than $0.22$.

This value of $\lambda$ will restrict the number of solutions
because first order calculations predict a dependency of the
Cabibbo angle on $\lambda$. If Yukawa matrices are given in terms
of powers of $\lambda$, so will, to leading order, the CKM matrix
\cite{fn}. The experimental uncertainty on the average value of
the Cabibbo angle \cite{lr,gh}
\begin{equation}
|{\bf V}_{12}|=0.2205\pm0.0018
\end{equation}
is very small. Keeping in mind that, as noted by Olechowski and
Pokorski \cite{op}, $|{\bf V}_{12}|$ is almost invariant (it
changes by less than 0.1\%) when renormalized from $M_W$ to
$M_{GUT}$, we will always try to keep close to
$\lambda\simeq0.22$.

In order to remedy the solution to eq.~(\ref{eqc3f}), we will
examine the assumptions that play a significant role since the
equation itself is robust.  It is robust because $C_3$ is related
to the exponent of a small parameter, so a small change in $C_3$
would change the determinants by orders of magnitude. With
$\lambda\simeq0.22$, we estimate that unless the order unity
factors in the Yukawa matrices all conspire to shift the balance in
one direction, they could increase $C_3$ by as much as two or
three. One should also note that there are no top or bottom Yukawa
couplings in eq.~(\ref{eqc318}), so the result is independent of
$\tan \beta$.

One assumption that affects the value of $C_3$ more significantly
is $f_\theta=1$. If, for example, $f_\theta$ were $0.78$, then the
expected and calculated values of $\lambda$ would be reconciled.

Another assumption that can be relaxed concerns the values of the
Kac-Moody levels. If, instead, $k_2=k_3=k_{GUT}=2$ while $k_{\rm
grav}=1$, then both eq.~(\ref{eqkgut}) relating $g_s$ and $g_{GUT}$
and eq.~(\ref{eqckratio}) relating $C_3$ and $C_{\rm grav}$ must
change. (Models with $k_{GUT}=2$ have been increasingly popular
with string theorists \cite{l,cc,af}.) The final relation for
$\lambda$, eq.~(\ref{eqc3l}), becomes
\begin{equation}
\label{eqc3lkap}
\lambda = f_\theta \sqrt{g_{GUT}\, C_3\over8\pi^2\sqrt{k_{GUT}}}
\simeq 0.079 f_\theta \sqrt{C_3}.
\end{equation}
Now, $\lambda\sim0.24$ when $C_3 = 9$.

Lastly, we note that $\lambda$ also depends on the value of $g_s$
by the above equation (for both $k_{GUT}=1$ and $k_{GUT}=2$).
However, we know that $f_\theta$ has a linear dependence on $g_s$,
so that
\begin{equation}
\label{eqfg}
\lambda\propto g_s^{3/2}.
\end{equation}
In order to attain a value of $\lambda\sim0.22$, $g_s$ would have
to be reduced from $0.7$ to $0.59$. This would require
$\alpha_{GUT}=1/36$, which is too low according to most models.

\section{Detailed Examples}
\label{secexamples}

We are now ready to examine in detail the $U(1)_X$ charge
assignments, which, subject to the constraints discussed in
sections~\ref{secanomaly} and \ref{secpq}, let us calculate fermion
mass matrices. We would then compare the quark and lepton masses
with relations~(\ref{eqratios}) and demand a phenomenologically
viable CKM matrix. Ideally, among all the possible charge
assignments we would find at least one that satisfies all the
constraints and predicts masses and mixings within experimental
bounds.

Although the fifteen charges of the quark and lepton fields may
seem like many free parameters, they are in fact overconstrained.
If we demand $\lambda=0.22$, equations (\ref{eqc3l}) and
(\ref{eqdet2}) become two independent predictions for~$C_3$. {\it A
priori\/} it is not obvious that the two numbers should even be of the
same order of magnitude. When $f_\theta$ and $k$ are taken into
account, in the context of a particular string compactification,
the two predictions will be more than just order of magnitude
estimates. If we do {\em not\/} require $\lambda=0.22$, then we
have to be able to produce the correct Cabibbo angle from a texture
given in terms of powers of the calculated $\lambda$.

Furthermore, the charges are integers, and they are constrained by
the mixed anomalies. One of the constraints is non-linear. It is
not guaranteed that there will be any solution, much less that it
will correspond to realistic masses and mixings.

\subsection{Positive Charges for Matter Fields}
\label{secpositive}

We begin the search for a detailed model by assuming that all the
standard model fields have nonnegative $U(1)_X$ charges. We also
assume, following Bin\'etruy and Ramond \cite{br}, that because of
a possible $\mu$ term, the Higgs doublets have zero $X$ charge. We,
therefore, form the Yukawa matrices
\begin{equation}
\label{eqYuk}
{\bf Y}_u = f_u \left(
\begin{array}{ccc}
\lambda^{(q_{Q1}+q_{u1})/|q_\theta|} &
\lambda^{(q_{Q1}+q_{u2})/|q_\theta|} &
\lambda^{(q_{Q1}+q_{u3})/|q_\theta|} \\
\lambda^{(q_{Q2}+q_{u1})/|q_\theta|} &
\lambda^{(q_{Q2}+q_{u2})/|q_\theta|} &
\lambda^{(q_{Q2}+q_{u3})/|q_\theta|} \\
\lambda^{(q_{Q3}+q_{u1})/|q_\theta|} &
\lambda^{(q_{Q3}+q_{u2})/|q_\theta|} &
\lambda^{(q_{Q3}+q_{u3})/|q_\theta|}
\end{array}
\right)
\end{equation}
for the up sector, and similarly for the down and lepton sectors.
Eq.~(\ref{eqYuk}) is an order of magnitude relationship, so that
each of the entries will be multiplied by a number of order unity.
The factors of order unity will be necessary to introduce CP
violation, which we are ignoring in this paper.

We would like to make $|q_\theta|$ the smallest charge unit,
i.e.\ $q_\theta=-1$ (in our convention all $X$ charges are
integers). That, however, does not lead to phenomenologically
acceptable mass matrices: every row of matrix (\ref{eqYuk}) is
equal to $\lambda^{\mbox{\scriptsize some power}}\times
(\mbox{some other row})$. Similarly, there is only one independent
column. The mass matrix, with two zero eigenvalues, does not
reproduce the observed mass hierarchy even qualitatively. Although
the factors of order unity multiplying the entries in ${\bf Y}_u$
will in general move $\;\det {\bf Y}_u\;$ away from zero, fermion
masses and mixings would then very strongly depend on those unknown
factors and not on the properties of the model we are trying to
investigate. Such a model may be realized in nature, however, we do
not know those factors. We will therefore set them to one in this
and the following section, and impose conditions to avoid zero
determinants. We will find that those conditions are too rigid to
produce a realistic example. Since they are not based on physical
principles, in section~\ref{sectf} we make an arbitrary but limited
choice of the ``texture factors'' to see how much can be achieved
by eliminating the artificial constraints.

Without texture factors, any matrix of the form
${\bf Y}_{ij}=\lambda^{a_i+b_j}$ will have rank one. A possible
solution is to use matrices with texture zeros:
$({\bf Y}_u)_{ij}=\lambda^{(q_{Qi}+q_{uj})/|q_\theta|}$ when
$(q_{Qi}+q_{uj})/|q_\theta|$ is a nonnegative integer, zero
otherwise.  If the sum of the charges $q_{Qi}+q_{uj}$ is not a
multiple of $|q_\theta|$, the corresponding term in the Lagrangian
is forbidden by charge conservation. If it is negative, it could
only be matched by a power of $\theta^\ast$. This is impossible in
supersymmetric theories because the superpotential must be
holomorphic in $\theta$.

With positive charges for the standard model fields, $q_\theta=-1$
does not allow texture zeros. We find that $q_\theta=-2$ is also
not enough.  The pattern of the texture zeros depends only on the
remainder from the division of $q_{Qi}+q_{uj}$ by $|q_\theta|$. Of
the three numbers $q_{u1}$, $q_{u2}$, $q_{u3}$, either at least two
will be even or at least two will be odd.  At least two columns of
${\bf Y}_u$, having the same pattern of texture zeros, will be
proportional. The result is a matrix of rank at most two.

In order to obtain a non-singular matrix we must have all $q_{Qi}$
different $(\mbox{mod }|q_\theta|)$ and all $q_{ui}$ different
$(\mbox{mod }|q_\theta|)$. That requires $|q_\theta|\ge3$. In
addition, one can see that no row and no column can have more than
one nonzero element. The only non-singular mass matrices we can get
in this model are rather sparse: they have six texture zeros.
Even without the assumption of left-right symmetry \cite{rr} there
are few possibilities:
\begin{eqnarray}
\label{eqmat}
\left(\begin{array}{ccc}
a & 0 & 0 \\ 0 & b & 0 \\ 0 & 0 & c
\end{array}\right),
\qquad
&
\left(\begin{array}{ccc}
0 & c & 0 \\ b & 0 & 0 \\ 0 & 0 & a
\end{array}\right),
\qquad
&
\left(\begin{array}{ccc}
a & 0 & 0 \\ 0 & 0 & b \\ 0 & c & 0
\end{array}\right),
\\
\left(\begin{array}{ccc}
0 & 0 & b \\ 0 & a & 0 \\ c & 0 & 0
\end{array}\right),
\qquad
&
\left(\begin{array}{ccc}
0 & a & 0 \\ 0 & 0 & b \\ c & 0 & 0
\end{array}\right),
\qquad
&
\left(\begin{array}{ccc}
0 & 0 & c \\ a & 0 & 0 \\ 0 & b & 0
\end{array}\right).
\nonumber
\end{eqnarray}
In addition, all the above textures can be reduced to a diagonal
form by a permutation of columns only. Permutations of columns of
a Yukawa mass matrix can be written as a multiplication on the
right by a unitary permutation matrix, and it is easy to show that
they do not change the resulting masses or mixings. (Permutations
of rows, which correspond to multiplication on the left, will not
change masses or mixings as long as they are done simultaneously in
the up and down sector.) We conclude that the masses and mixings
will be the same as if all mass matrices were diagonal, i.e.\ there
will be no flavor mixing.

Can we make things better by giving up some of the assumptions? In
the following section we will let all particles have charges of
either sign. Here we only note that if we allow the Higgs doublet
to have a positive charge (at the expense of the $\mu$ term), we
are merely shifting the charges of the up, down or lepton sector
without changing any of the conclusions. If the Higgs charge is a
multiple of $|q_\theta|$, the textures do not change. Otherwise,
the texture zeros will change their positions, but we will still
have exactly one nonzero entry in each row and each column.

It should be stressed again that our failure to find a workable
example of a model with only positive charges for the standard
model fields does not in any way rule out such models.  Our
conclusion here is that we cannot obtain an acceptable mass matrix
to study unless we know the exact factors of order unity in front of
the powers of $\lambda$ in (\ref{eqYuk}). It would be interesting
to come back to this exercise when we can make an informed choice
of those factors.

\subsection{Allowing Negative Charges}
\label{secnegative}

We now turn to the analysis of the model in which the matter fields
are allowed to have negative as well as positive charges. That will
be a source of texture zeros, and give us more flexibility in
constructing the mass matrices.

We need to find the conditions necessary to obtain a non-singular
matrix, and start by reordering the quarks and leptons so that
$q_{Q1} \ge q_{Q2} \ge q_{Q3}$,
$q_{u1} \ge q_{u2} \ge q_{u3}$, etc.
This is the same as permuting the rows and columns of the mass
matrix; it can only change the determinant of {\bf Y} by a sign,
and will not change the masses or mixings.  Now in
eq.~(\ref{eqYuk}) we put a texture zero wherever $q_{Qi}+q_{uj}<0$.
Keeping in mind that no two columns and no two rows can have the
same pattern of texture zeros, we are left with
\begin{equation}
\label{eqYukneg}
{\bf Y}_u = f_u \left(
\begin{array}{ccc}
\lambda^{q_{Q1}+q_{u1}} &
\lambda^{q_{Q1}+q_{u2}} &
\lambda^{q_{Q1}+q_{u3}} \\
\lambda^{q_{Q2}+q_{u1}} &
\lambda^{q_{Q2}+q_{u2}} &
0 \\
\lambda^{q_{Q3}+q_{u1}} &
0 &
0
\end{array}
\right).
\end{equation}
We are assuming $q_\theta=-1$ in this section.  The texture
(\ref{eqYukneg}) is rich enough to ask whether it is possible to
obtain realistic fermion masses and mixings.

To answer this question, we have done a computerized search by
trying out all possible charge assignments for the quarks and
leptons (in a range from $-10$ to 10), imposing the anomaly
constraints, calculating the fermion masses and aiming to be as
close to the ratios (\ref{eqratios}) as possible. (We were limited
in how close we could get by the relations (\ref{eqc3l}) and
(\ref{eqdet2}) between the determinants of the quark mass matrices
and the hierarchy parameter $\lambda$.) For those charge sets that
produced the best fermion masses, we then computed the CKM matrix.
While none of the results reproduce experimental data, we have
found many that were not unreasonable. Below we give an example.
For $g_s=0.7$ and the following $X$ charges,
$$
\begin{tabular}{c|rrrrr}
$i$&  $q_{Qi}$  &  $q_{ui}$  &  $q_{di}$  &  $q_{Li}$  &  $q_{ei}$  \\
\hline
1  &    9    &    9    &    7    &    1    &   10    \\
2  &    7    &    2    &  $-4$   &  $-7$   &    9    \\
3  &  $-4$   &  $-9$   &  $-9$   & $-10$   &    3
\end{tabular}
$$
we have $C_3=10$ and $\lambda=0.30$,
which results in the fermion mass ratios
\begin{eqnarray*}
\rule[-2pt]{0pt}{14pt}
{m_u \over m_t}      =   1.8\times10^{-5} \;, \qquad\quad
{m_d \over m_b}     &=&  2.6\times10^{-2} \;, \qquad\quad
{m_e \over m_\tau}   =   7.9\times10^{-3} \;, \\
\rule{0pt}{18pt}
{m_c \over m_t}      =   2.3\times10^{-2} \;, \qquad\quad
{m_s \over m_b}     &=&  2.6\times10^{-2} \;, \qquad\quad
{m_\mu \over m_\tau} =   8.9\times10^{-2} \;, \\
\rule[2pt]{0pt}{16pt}
{m_t \over f_u}      =   1.0              \;, \qquad\qquad\qquad\;
{m_b \over f_d}     &=&  1.0              \;, \qquad\qquad\qquad\;
{m_\tau \over f_d}   =   1.0              \;,
\end{eqnarray*}
and the CKM matrix
\begin{equation}
\label{eqV}
{\bf V} = \left(
\begin{array}{ccc}
 0.96             & 0.27              & 6.2\times10^{-5}  \\
-0.27             & 0.96              & 2.0\times10^{-10} \\
-6.0\times10^{-5} & -1.6\times10^{-5} & 1.0
\end{array}
\right).
\end{equation}
The 90\% confidence experimental limits on the magnitude of the CKM
matrix elements \cite{rev}, renormalized to the GUT scale
\cite{op}, are
\begin{equation}
\left(
\begin{array}{ccc}
0.9747\mbox{ to } 0.9759 &
0.218 \mbox{ to } 0.224  &
0.001 \mbox{ to } 0.003  \\
0.218 \mbox{ to } 0.224  &
0.9738\mbox{ to } 0.9752 &
0.021 \mbox{ to } 0.032  \\
0.002 \mbox{ to } 0.010  &
0.020 \mbox{ to } 0.032  &
0.9995\mbox{ to } 0.9998
\end{array}
\right).
\end{equation}

One obvious defect of the above example is that $\lambda$ and the
Cabibbo angle ${\bf V}_{12}$ are too big, as discussed in
section~\ref{secpq}. Another is the degeneracy of the $s$ and $d$
quarks. Again, this is a generic feature of the examples based on
the texture~(\ref{eqYukneg}). The reason is that the dominant terms
in (\ref{eqYukneg}) are on the antidiagonal; the terms in the upper
left triangle are orders of magnitude smaller. Therefore,
(\ref{eqYukneg}) can be thought of as an antidiagonal matrix with a
very small perturbation. Since a permutation of columns can cast it
in an almost diagonal form, such a matrix will not lead to any
appreciable flavor mixing {\em unless\/} two of the eigenvalues of
$\bf Y\,Y^\dagger$ are degenerate. In such a case the choice of the
basis in the eigenspace is arbitrary, and it is easy to obtain
large mixing angles. We were looking for ${\bf V}_{12} \simeq
0.22$, so it is understandable for all the examples to have
degenerate quark masses.

The last thing we noticed is that the ``small'' entries (1,3),
(3,1), (2,3), (3,2) of the CKM matrix~(\ref{eqV}) are much smaller
than what we know from the experiment (even taking renormalization
into account). The small mixing in the heavy flavor sector can be
understood by noting that the heavy quarks are not degenerate in
mass, and the magnitude of the mixing in this case is determined by
the magnitude of the off-diagonal perturbation.

\subsection{Allowing a Texture Factor}
\label{sectf}

Until now, we have been trying to keep all the factors of order unity
in the Yukawa mass matrices equal to one. We tried to avoid
the problem of singular mass matrices by restricting our search to
matrices with enough texture zeros to be nonsingular. In this
section, we want to make the matrices non-singular by introducing
coefficients different from unity.

To avoid introducing 27 free parameters, we make a rather arbitrary
choice of the coefficients: we introduce one parameter, the
``texture factor'' (TF), which will multiply the (2,3), (3,2) and
(3,3) elements of ${\bf Y}_u$ and ${\bf Y}_e$. The corresponding
elements of ${\bf Y}_d$ are divided by TF. That is the minimal
intervention needed to make the determinants nonzero in most
cases. We chose to modify the entries in the heavy quark sector so
that the predicted Cabibbo angle would not depend strongly on TF.
We decided to divide, rather than multiply, in ${\bf Y}_d$, to make
the determinant relations such as~(\ref{eqdet1}) minimally
sensitive to TF. Finally, we considered only nine discrete values
of TF: $-1$, $\pm2$, $\pm\sqrt{2}$, $\pm1/2$, $\pm1/\sqrt{2}$.
With negative as well as positive charges allowed, but without
requiring any particular pattern of texture zeroes, we were able to
find better examples than before.  With $g_s=0.7$ and the $X$
charges
$$
\begin{tabular}{c|rrrrr}
$i$  &  $q_{Qi}$  &  $q_{ui}$  &  $q_{di}$  &  $q_{Li}$  &  $q_{ei}$  \\
\hline
1 &    5    &    3    &    2    &    6    &    5    \\
2 &    4    &    2    &  $-1$   &  $-2$   &    4    \\
3 &  $-2$   &    1    &  $-1$   &  $-5$   &  $-1$
\end{tabular}
$$
we have $C_3=10$ and $\lambda=0.30$.
With the texture factor ${\rm TF}=+2$,
we find the fermion mass ratios
\pagebreak[3]
\begin{eqnarray*}
\rule[-2pt]{0pt}{14pt}
{m_u \over m_t}      =   7.5\times10^{-6} \;, \qquad\quad
{m_d \over m_b}     &=&  3.3\times10^{-3} \;, \qquad\quad
{m_e \over m_\tau}   =   2.3\times10^{-3} \;,\\
\rule{0pt}{18pt}
{m_c \over m_t}      =   2.3\times10^{-3} \;, \qquad\quad
{m_s \over m_b}     &=&  3.1\times10^{-2} \;, \qquad\quad
{m_\mu \over m_\tau} =   8.7\times10^{-2} \;,\\
\rule[2pt]{0pt}{16pt}
{m_t \over f_u}      =   2.0              \;, \qquad\qquad\qquad\;
{m_b \over f_d}     &=&  1.0              \;, \qquad\qquad\qquad\;
{m_\tau \over f_d}   =   1.0              \;,
\end{eqnarray*}
and the CKM matrix
\begin{equation}
\label{eqV2}
{\bf V} = \left(
\begin{array}{ccc}
  0.98             & 0.20              & 5.0\times10^{-5} \\
 -0.20             & 0.98              & 3.5\times10^{-4} \\
  2.1\times10^{-5} & -3.6\times10^{-4} & 1.0
\end{array}
\right).
\end{equation}

Although the small elements of the CKM matrix are still about an
order of magnitude too small, the above example has no other
obvious defects. It reproduces the mass ratios~(\ref{eqratios})
and the Cabibbo angle fairly well, the bottom quark and the tau
have equal masses at the unification scale, and the top quark is
the heaviest.  An even better fit (including the small elements of
the CKM matrix) can be obtained for Kac-Moody level $k_{GUT}=2$,
thanks to a much better agreement between (\ref{eqc318}) and
(\ref{eqc3l}). This shows that, finally, it is possible to obtain
realistic masses and mixings as a result of an anomalous $U(1)_X$
family symmetry.

\section{Conclusions}
\label{secconclusions}

In this paper, we have examined the idea that the hierarchy
parameter in Froggatt-Nielsen type models is related to
a spontaneously broken anomalous $U(1)_X$ gauge symmetry left
over from string compactification. Given the details of a string
compactification, the condition of preserving supersymmetry
(\ref{eqsqrt}) predicts the value of $\lambda$, hence the
complete mass matrices, fermion masses and mixings.

Without a complete string model at hand, we looked for
model-indepen\-dent features. We found a strong constraint from
anomaly cancellation (\ref{eqcratio}) that gives $\lambda$ in
terms of the quark $X$ charges only. There is another constraint
on those charges (\ref{eqc318}) from the known value of the
product of all quark masses. Assuming $\lambda=0.22$, the two
independent predictions (\ref{eqc3l}) and (\ref{eqc39}) agree as
an order of magnitude relation. Their agreement in a definite
model depends on $f_\theta$, the coefficient that enters
$\lambda = f_\theta\langle\theta\rangle/M$, and on the Kac-Moody
level of the gauge group, $k_{GUT}$.

We also wanted to verify that it is possible to obtain realistic
masses and mixings in this framework. We examine integer charge
assignments satisfying the anomaly cancellation constraints, and,
with the $\lambda$ determined by those charges, find some promising
examples. It would be very interesting to do the same with a
definite string model. The values for $f_\theta$, $k_{GUT}$ and the
order unity texture factors in mass matrices would then be
specified, leaving no free parameters. The method presented in this
paper gives us a powerful tool to narrow down the set of possible
string compactifications. We think that there will be only a small
number (if any) of models compatible with the idea of predicting
$\lambda$ from $U(1)_X$ charges.

One desirable feature of the CKM matrix we have not addressed in
this paper is CP violation. It is possible to introduce CP
violation by assuming that the order unity texture factors are
complex \cite{fn}, but that generically leads to large CP
violation. Another way is to use a model with two $U(1)_X$
breaking fields, $\theta_1$ and $\theta_2$. The phase of the VEV
of a single complex $\theta$ field can always be rotated away by
a gauge transformation, but for two complex fields there is a
gauge-invariant phase difference, so that in general we can make
only one of them (say $\langle\theta_1\rangle$) real.

The powers of $\langle\theta_1\rangle$ and $\langle\theta_2\rangle$
will give imaginary parts to the mass matrix elements. If
$\langle\theta_2\rangle \ll \langle\theta_1\rangle$, then the
imaginary parts will be necessarily small, leading to naturally
small CP violation. Work on the details of the two-theta model
is in progress.

\section*{Acknowledgements}

We have the pleasure to thank Jeff Harvey for suggesting the idea
of this paper, and for many useful discussions until its
completion. We would also like to acknowledge conversations with
Tony Gherghetta, Aaron Grant, Erich Poppitz and Mihir Worah.
This research was supported in part by NSF Grant No.\ PHY-9123780.

\end{document}